\documentclass[11pt,fleqn]{article} 
\usepackage{amssymb}

\usepackage{amsmath}

\usepackage{amsmath,amssymb,amsthm}

\begin{document} {\Large \begin{center} {\Large \textbf{Comment to "General Solution
to Unidimensional Hamilton-Jacobi Equation"(arXiv:1302.0591v1)}}

\vskip 20pt {\large \textbf{Irina YEHORCHENKO}}

\vskip 20pt {Institute of Mathematics of NAS Ukraine, 3 Tereshchenkivs'ka Str., 01601
Kyiv-4, Ukraine} E-mail: iyegorch@imath.kiev.ua \end{center}

\vskip 50pt
\begin{abstract} We present previous results on the general solution of
the multidimensional Hamilton-Jacobi equation $\frac{\partial u}{\partial t} - \frac{\partial
u}{\partial x_a} \frac{\partial u}{\partial x_a}= 0$ and methods that were used to
find such general solution.
\end{abstract}

\section{Introduction}

The paper \cite{MLE} says that {\it "Unfortunately, due to the nonlinearity of this
equation }(the Hamilton Jacobi equation - IY) {\it till now there is no available technique to determine a general solution"}. 

I would like to point out that such technique does
exist (the method of the hodograph transformation), and it was used by us in 1991
\cite {FYe Schr} (following \cite{ZhdanovRevenko}) to find the general solution for
the Hamilton-Jacobi equation

\begin{equation} \label{HE}
\frac{\partial u}{\partial t} - \frac{\partial u}{\partial x_a} \frac{\partial u}{\partial x_a}= 0.
\end{equation}

Here $a=1, ..., n$ (so, we deal with the arbitrary number of the space variables).

\section{General solution of the Hamilton-Jacobi equation}

Finding of the general solution of the Hamilton-Jacobi equation in \cite{FYe Schr}
was regarded as an intermediate task and was not specifically acknowledged, so it
might have been overlooked. However, it is a technical problem, and I really could not
claim that our paper \cite{FYe Schr} was the first instance when such general
solution was mentioned. General solution of the Hamilton-Jacobi equation in a different form was considered in \cite{Epstein}.

Anyway, such general solution of the rank $n$ 

(that means that
$\det\{\frac{\partial u^2}{\partial x_a \partial x_b}\}\ne 0$)
has the form

\begin{equation}
 u=x_a y_a - \frac{t}{2} y_a y_a + \Phi(\overrightarrow{y}),
\label{sol}
\end{equation}

\begin{equation}
  x_a - ty_a +\frac{\partial\Phi}{\partial y_a} = 0, \label{cond}
\end{equation}

$y=(y_1, ..., y_n)$ are parameter functions, $\Phi$ is an arbitrary function of its
parameters subject to the condition (\ref{cond}). We assume summation from $1$ to $n$
by the repeated indices.

Rank of the solution is the rank of the matrix
$\{\frac{\partial u^2}{\partial x_a \partial x_b}\}$.

\section{Hodograph transformation and its use for finding of solutions of PDE} The
method of the hodograph transformation (interchanging of some of the dependent and
independent variables) in the equation was described e.g. in \cite{CF}.

The hodograph transformations for the equation (\ref{HE}) 

and ${\rm rank}\{\frac{\partial u^2}{\partial x_a \partial x_b}\}=n$
is as follows:

\begin{equation} \label{ht}
y_0=t, y_a=\frac{\partial u}{\partial x_a},
\end{equation}
\begin{equation}
H(y)= x_a\frac{\partial u}{\partial x_a} - u,\nonumber
\end{equation}

Then we come to the Hamilton-Jacobi equation in new variables

\begin{equation}
\frac{\partial H}{\partial y_0}= -y_a y_a. \nonumber
\end{equation}

The general solution in terms of the function $H$:
\begin{equation} \label{H}
H= - t y_a y_a - \Phi(\overrightarrow{y}).
\end{equation}

Using to the expression (\ref{H}) the inverse hodograph transformations (\ref{ht}),
we will receive the solution (\ref{sol}) with the conditions (\ref{cond}).

We considered the case of solutions of the rank $n$; the procedure for the ranks from
0 to $n-1$ is similar with the relevant number of the parameter functions.

The hodograph transformations were used in \cite{ZhdanovRevenko} for finding of the general solution of the system of eikonal and d'Alembert equations.

As to utilisation of the hodograph transformations for solving of other PDEs and
systems of PDEs, see \cite{Meleshko-book}, \cite{Kevorkian}.

The method presented in \cite{MLE} is technically the same as the hodograph
transformation, but in different terminology, and gives the same result in different notations formula (5) in \cite{MLE}).

The hodograph transformations is a straightforward and powerful technique to
linearise and solve nonlinear PDEs (not only first-order), and probably worth writing of a detailed tutorial.

 } 
\end{document}